\documentclass[11pt, draftclsnofoot, onecolumn]{IEEEtran}
\usepackage[margin=1in,right=1.25in,left=1.25in]{geometry}

\usepackage{tabu}
\usepackage{soul}
\usepackage[caption=false,font=footnotesize]{subfig}
\usepackage{url}

\usepackage{amsfonts}
\usepackage{amsthm}
\usepackage{amssymb}
\usepackage{amsmath}
\usepackage{mathtools}
\usepackage{color}  
\usepackage{multicol} 
\newcommand{\ignore}[1]{}
\usepackage{graphicx}

\begin{document}

\title{Low-Power Neuromorphic Hardware for Signal Processing Applications}
\author{
Bipin Rajendran$^1$, Abu Sebastian$^2$, Michael Schmuker$^3$, Narayan Srinivasa$^4$, and Evangelos Eleftheriou$^2$\footnote{ 
$^1$New Jersey Institute of Technology, $^2$IBM Research Zurich, $^3$University of Hertfordshire, $^4$Intel Labs.\\
\indent Corresponding author: bipin@njit.edu}
\vspace{-.6in}
}

\maketitle

 \begin{abstract} 
Machine learning has emerged as the dominant tool for implementing complex cognitive tasks that require supervised, unsupervised, and reinforcement learning. While the resulting machines have demonstrated in some cases even super-human performance, their energy consumption has often proved to be prohibitive in the absence of costly super-computers. Most   state-of-the-art machine learning solutions are based on memory-less models of neurons. This is unlike the neurons in the human brain that encode and process information using temporal information in spike events. The different computing principles underlying biological neurons and how they combine together to efficiently process information  is believed to be a key factor behind their superior efficiency compared to current machine learning systems. Inspired by the time-encoding mechanism used by the brain, third generation spiking neural networks (SNNs) are being studied for building a new class of information processing engines.

Modern computing systems based on the von Neumann architecture, however, are ill-suited for efficiently implementing  SNNs, since their performance is limited by the need to constantly shuttle data between physically separated logic and memory units. Hence, novel computational architectures that address the von Neumann bottleneck are necessary in order to build systems that can implement SNNs with low energy budgets.  In this paper, we  review some of the architectural and system level design aspects involved in developing a new class of brain-inspired information processing engines that mimic the time-based information encoding and processing aspects of the brain. \end{abstract}

\vspace{-.2in}
\section{Introduction}

While machine learning algorithms based on deep neural networks (DNNs) have demonstrated human-like or even super-human performance in several complex cognitive tasks, a significant gap exists between the energy and efficiency of the computational systems that implement these algorithms compared to the human brain. Most of these algorithms run on conventional computing systems such as central processing units (CPUs), graphical processing units (GPUs) and field programmable gate arrays (FPGAs). Recently, digital or mixed-signal application specific integrated circuits (ASICs) have also been developed for machine learning acceleration. However, as Moore's law scaling is coming to an imminent end, the performance and power efficiency gains from technology scaling of these conventional approaches are diminishing. Thus, there are significant research efforts worldwide in developing a profoundly different approach to computing for artificial intelligence (AI) applications inspired by biological principles. 

In the traditional von Neumann architecture, a powerful processing unit operates sequentially on data fetched from memory. In such machines, the von Neumann bottleneck is defined as the limitation on performance arising from the ``chokepoint'' between computation and data storage. Hence, the research focus has been not only on designing new AI algorithms, device technologies, integration schemes, and architectures but also on overcoming the CPU/memory bottleneck in conventional computers. Spiking neural networks (SNNs) are the third generation of artificial neuron models that leverage the key time-based information encoding and processing aspects of the brain. Neuromorphic computing platforms aim to efficiently emulate SNNs in hardware by distributing both computation and memory among large number of relatively simple computation units, viz. the neurons, each passing information, via asynchronous spikes, to hundreds or thousands of other neurons through synapses  \cite{Mead-02}. The event-driven characteristics of SNNs have led to highly efficient computing architectures with collocated memory and processing units, significantly increased parallelism and reduced energy budgets. Such architectures have been demonstrated in neuromorphic implementations such as SpiNNaker from the University of Manchester \cite{SpiNNaker}, IBM's TrueNorth \cite{truenorth}, Intel's Loihi \cite{loihi},  BrainScaleS built by Heidelberg University \cite{Meier2015}, NeuroGrid from Stanford \cite{Y2014benjaminProcIEEE}, DYNAP from INI Zurich \cite{Moradi2018}, and ODIN from Catholic University Louvain \cite{Frenkel2019}. Moreover, breakthroughs in the area of nanoscale memristive devices have enabled further improvements in area and energy efficiency of mixed digital-analog implementations of synapses and spiking neurons \cite{Kuzum13,Y2016tumaNatNano}. 

In this paper, we provide a high-level description of the design objectives and approaches that are currently being pursued for building energy efficient neuromorphic computing platforms. It is organized as follows: Section \ref{info} provides a brief overview of the information processing aspects of the human brain, which have motivated the design of artificial neuromorphic computing platforms. Section \ref{blocks} describes the objectives and specifications of the building blocks of neuromorphic platforms - hardware neurons,  synapses, and architectures to provide connectivity between them. Section \ref{design} introduces several design choices and principles underlying non-von Neumann computing architectures. Section \ref{chips}  covers the salient aspects of some of the state-of-the-art neuromorphic chips. Section \ref{devices} provides an overview of some of the research into developing computational memory units based on nanoscale memristive devices and how they can execute certain computations in place avoiding the von Neumann bottleneck and also naturally capture timing-based correlations in signals. In section \ref{apps}, we discuss some signal processing applications of these neuromorphic computing platforms and finally conclude the paper in Section \ref{Challenges} by reviewing the future outlook and important challenges for the field.

\section{Information processing in the brain}
\label{info}

``Computing'' in the brain follows a completely different paradigm than today's conventional computing systems. Whereas conventional systems are optimized to transmit and modify numerical representations of data, the brain operates on timed events called action potentials or \emph {spikes}. Neurons receive these spikes via synapses, which convert them into small changes in the cell's membrane potential. The neuron integrates these changes in potential over time, and under certain conditions, for instance, when many spikes arrive within a short time, the neuron emits a spike. Spikes can be considered messages in the computing sense, except that they carry no information other than their time of generation and their source. Computing in the brain can thus be described as fully event-driven, non-blocking and imperatively concurrent, with very lightweight compute nodes (the neurons), that communicate via tiny messages, the spikes. 

There are about $10^{11}$ neurons in the human brain, and it {is} estimated that there are about $5,000-10,000$  synapses per neuron in the human neocortex. Thus, connectivity is sparse, with a neuron receiving input from  $10^{-6} \%$ of all other neurons (probably even less, considering that an axon may form multiple synapses on the same dendrite). At the same time, the total number of connections is huge, in the order of $10^{15}$. This is vastly different from the low fan-out connectivity that is common in conventional computers.

\subsection{Signal encoding in the brain}
The event-driven nature of computing also applies to stimulus encoding by the sensory organs. Generally, sensory coding emphasizes \emph{changes} in the stimulus, rather than accurate coding of constant levels. These changes can be considered ``events'' that are translated into spikes for downstream processing. For example, ganglion cells in the retina transmit a spike if the change in local contrast in their receptive field exceeds a threshold. They then adapt to the new level of local contrast. More spikes are produced only when local contrast rises further. This encoding scheme has three advantages over uniformly-spaced signal sampling in conventional signal processing: first, it produces a sparse code that transmits information only when the input signal is changing; second, it is not limited by a fixed sample rate in the maximum frequency it can encode;  third, it can lead to extremely reactive systems since the latency for feature extraction is not limited by the time between two samples, but only the minimum delay between two events, which is usually much shorter.   In neuromorphic devices, this encoding scheme has been implemented via a set of thresholds that trigger events to be fired upon the signal crossing them (Fig. \ref{fig:brainComp}, right top~\cite{LiuBook15}).

\subsection{``Learning'' in the brain}
\begin{figure*}[!ht]
	\centering
	\includegraphics[width=0.7\linewidth]{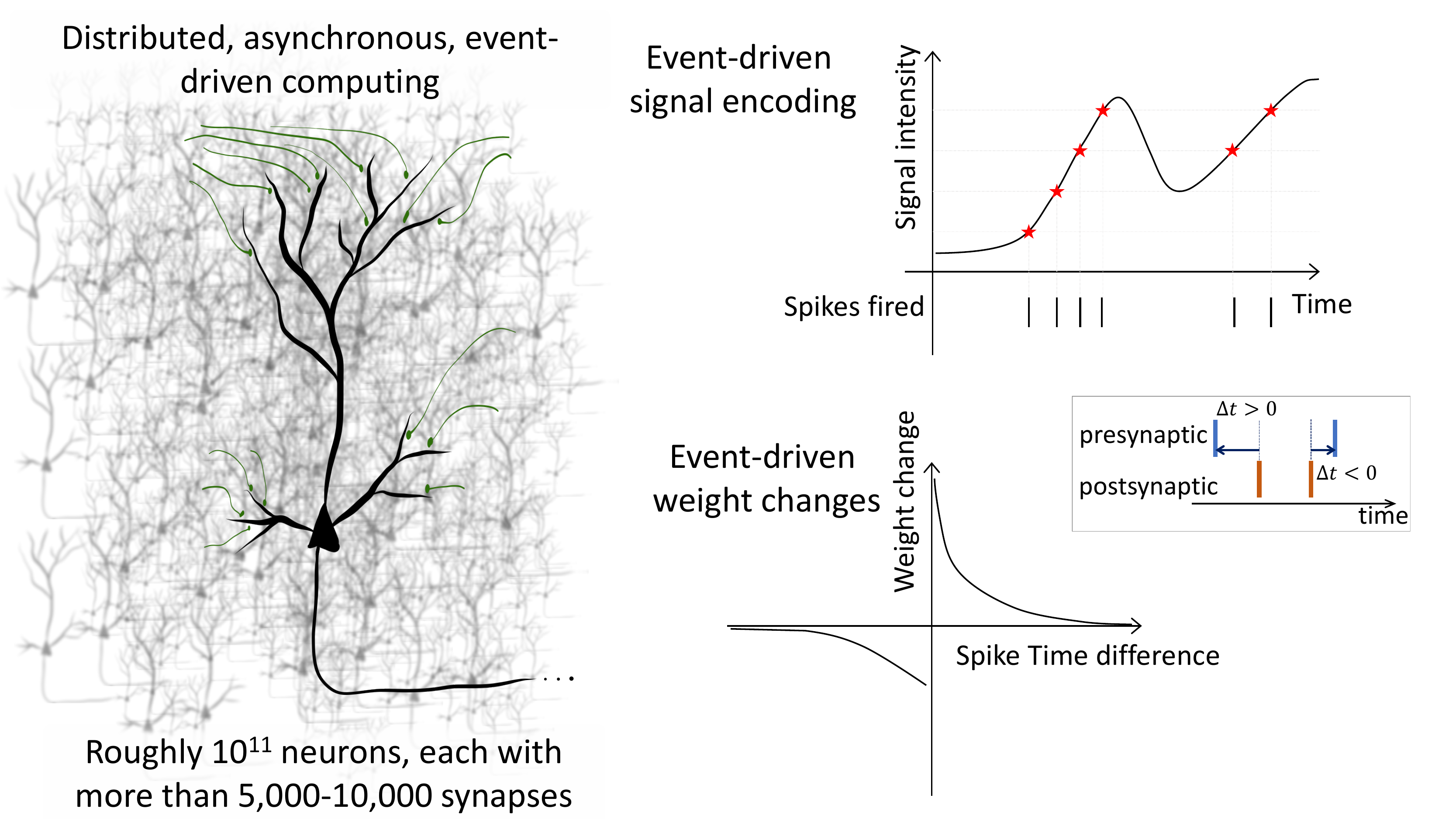}
	\caption{The brain employs a massively parallel computational network comprising of $\sim10^{11}$ neurons and $\sim10^{15}$ synapses (left) using event-driven signal encoding (top right) and weight updates (bottom right). For details, see text.
	}
	\label{fig:brainComp}
\end{figure*}

The connections in the brain are not fixed, but can change as a function of their activity; this process is often called ``learning''. Fundamental principles of synaptic changes have been uncovered that depend on the timing of spikes fired in the pre- and postsynaptic cells, thus termed Spike-Timing Dependent Plasticity or STDP \cite{BiPoo}. STDP implements a form of Hebbian learning, where a synapse gets strengthened (\emph{potentiated} in Neuroscience-lingo) if the presynaptic spike arrives within a certain time window preceding the spike of the post-synaptic cell, or weakened (\emph{depressed}) if the temporal order is reversed (Fig. \ref{fig:brainComp}, right bottom). Pre- and post-synaptic timing are thus two \emph{factors} that determine the change of a synaptic weight. 

In biology, the amount and direction of weight change are often influenced by neuromodulators such as Dopamine or Noradrenaline  which get released as a function of reward received by the organism.  Neuromodulators are thus a third factor in models of synaptic learning. They allow for the construction of powerful spiking learning rules that, for example, implement reinforcement learning \cite{Urbanczik:2009cc}. Taken together, the brain's massively parallel, event-driven computing combined with its extraordinary connectivity probably is the basis for its extremely high efficiency in signal processing, inference, and control. Furthermore, the action potentials used for communication as well as the synaptic currents in the brain have much smaller magnitudes than that of the electrical signals in silicon computers, as signals in the brain are encoded by the flow of a relatively smaller number of charge carriers (such as Na, K and Ca ions) through ion channels with highly selective, stochastic, and non-linear conductance characteristics. As a result of all these contributing factors, the brain is estimated to consume about $20\,$W even during demanding tasks like mastering a multi-player online computer game, whereas a conventional platform has been reported to use about 128,000 CPUs and 256 GPUs to achieve competitive performance (see \url{https://blog.openai.com/openai-five}).

This fundamental difference in computing architecture implies that porting algorithms from conventional machines to spike-based approaches can only have limited success. New algorithms are required that embrace the fundamentally event-based nature of sensing, learning, and inference in the brain, to leverage the full potential of spiking networks accelerated by neuromorphic hardware.

\section{Building blocks of neuromorphic systems}\label{blocks}
 
Although biological neurons and synapses exhibit a wide variety of complex dynamical behaviors, most hardware designs need to mimic only the key aspects that are essential for computation. At a high level, this includes the integrate-and-fire (I\&F) dynamics of neurons and spike-triggered communication and plasticity of synapses, which we describe below.

The central aspects of integrate-and-fire dynamics of neurons are described by the Hodgkin-Huxley equations which incorporate the voltage-dependent dynamics of sodium, potassium, and leak ion channels to determine the evolution of the membrane potential. While biological neurons exhibit more complex behaviors such as burst firing, chirping, post-inhibitory rebound, and spike-frequency adaptation, and though the computational significance of these has not been clearly established so far, there have been example hardware designs that mimic some of these behaviors using both digital CMOS and sub-threshold analog CMOS circuits. Model order reduction techniques have been used to reduce the complexity of the dynamical equations that describe these behaviors; some of the notable examples being the second order Izhikevich model, the adaptive exponential I\&F model, and the linear leaky I\&F (LIF) model. These are simpler to implement in hardware and are more commonly used in large scale neuromorphic designs. We describe the LIF model here as it is the most commonly used spiking neuron model. The membrane potential $V(t)$ evolves according to the  differential equation:
\begin{align}
    C \frac{dV(t)}{dt} &= -g_{L}(V(t)-E_L)+I_{syn}(t),&&
V(t)\leftarrow E_L, \text{  if  }  V(t)>V_T
	\label{eq:lif} 
\end{align}
 i.e., the input synaptic current $I_{syn}(t)$ is integrated across a leaky capacitor until the voltage exceeds a threshold   $V_T$, when a spike is issued and the membrane potential is reset to its resting value $E_L$. $C$ and $g_L$  model the membrane's capacitance and leak conductance, respectively.  The refractory period seen in biological neurons can be implemented by holding the membrane potential at $V(t)=E_L$, preventing current integration during that period. 

Synaptic communication is triggered by the arrival of a spike, causing a current to flow into downstream neurons. This is also a very complex process in the brain, involving the release of neurotransmitter molecules at the axon terminal, which then diffuses across the synaptic cleft, and binds with receptor molecules at the dendrite causing ionic currents to flow into the downstream neurons. These aspects are not modeled in most hardware implementations. Instead, the current through a  synapse with strength $w$ is calculated as 
\begin{align}
\scriptsize 
I_{syn} (t)   = w \times \sum_{{i}} \alpha( t-t^i) 
\label{eq:Isyn}
\end{align}
where $t^i$ represents  the time of arrival of spikes at the axon terminal and $\alpha(t)$ is a synaptic current kernel; $\alpha(t)=(e^{-t/\tau_1}-e^{-t/\tau_2})$,   $\alpha(t)=(t/\tau)e^{-t/\tau}$, and $\alpha(t)=\tau\delta(t)$ are some commonly used synaptic kernels. Note that this form of synaptic transmission also assumes that the current is independent of the post-synaptic neuron's potential, unlike in biology. Finally, in order to implement synaptic plasticity, the weight $w$ is updated based on learning rules  which are implemented in peripheral circuits, again in a spike-triggered fashion.

While the hardware design of neuronal and synaptic circuits involves a variety of trade-offs in area, power, reliability, and performance, the more challenging aspect of neuromorphic systems is supporting arbitrary connectivity patterns between spiking neurons. Since the computational advantage of neural networks emerges from the high fan-out, yet sparse, connectivity of neurons, hardware designs also need to support this crucial aspect. Furthermore, several forms of structural plasticity are observed in the brain where neurons can form (or remove) new synaptic connections based on activity. In the following section, we describe some of the architectural design choices that enable the realization of some of these aspects in silicon substrates.  

\section{System  design principles and approaches}\label{design}

\begin{figure*}[!ht]
	\centering
	\includegraphics[width=0.5\linewidth]{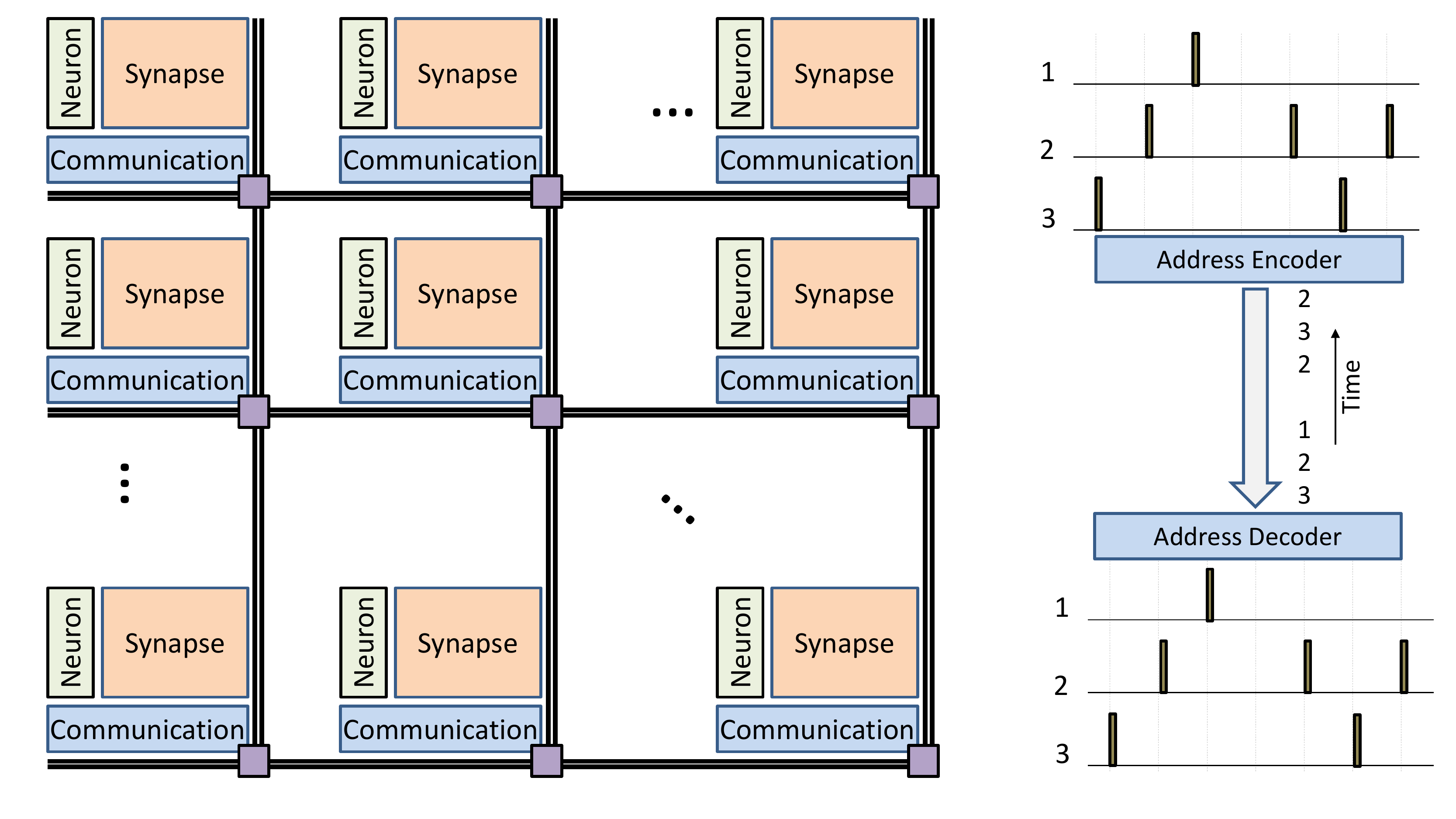}
	\caption{Architecture and communication protocols used in neuromorphic systems - a tiled array of neuromorphic cores with each core integrating neurons and synapses locally. Spikes are communicated between cores through a routing network using address event representation (AER) protocols. Adapted from \cite{truenorth, LiuBook15}.  }
	\label{fig:arch}
\end{figure*}

Collocation of memory and computation mitigates the von Neumann bottleneck in neuromorphic processors. Thus, inspired by neuroscience, the state-of-the-art architectural framework of SNN accelerators or co-processors comprises a network of neurosynaptic cores \cite{truenorth,loihi} that can efficiently implement scale-out neural networks as shown in Fig. \ref{fig:arch}. Each core consists of a crossbar array with electronic synapses at each cross-point and peripheral circuitry including local SRAM-based look-up tables for routing information and storage of local data. The peripheral circuitry implements the neuronal functionality (typically the LIF neuron), the read/write electronics as well as the driver circuits that control the voltages on the input (axon, horizontal lines) and output (dendrites, vertical lines) wires. Thus, a neurosynaptic core represents a layer of the spiking neural network and the passing of information between layers (cores) is enabled by a time-multiplexed communication fabric.

Since neuron spiking rates are orders of magnitude slower than digital electronics and jitter and delay through digital electronics (propagation and transition delay) is insignificant compared to axonal delays and neuron time constants, the networks typically used to multiplex information from one neurosynaptic core to another are packet-switched networks, using the so-called address event representation (AER) protocol. In this scheme, each neuron has a unique address which is transmitted to the destination axons when it spikes;  the time of spiking is hence encoded implicitly (Fig. \ref{fig:arch}).  Note that the AER protocol also allows efficient and flexible off-chip interconnect for large-scale multi-chip network platforms. The overall architecture is brain-inspired in the sense that the neurosynaptic core, via its crossbar array, mimics the dense local connectivity of neurons within the same layer, whereas the network interconnect topology with the AER protocol allows sparse long-range connections. From a pragmatic hardware-implementation-oriented viewpoint, such an SNN accelerator architecture with the appropriate core-to-core communications protocol is reminiscent of a dataflow engine, i.e., data flow from core-to-core in an asynchronous fashion. See  \cite{LiuBook15} for a general review of neuromorphic design principles. 
 
Although the core-to-core interconnect fabric is implemented with digital CMOS logic, the neurosynaptic core itself can be implemented by using analog/mixed-signal, digital, or memristive technologies.  In its simplest form, the neuron membrane potential can be implemented in the analog domain as a voltage across a capacitor or as a multi-bit variable stored in digital latches.  The analog design approaches have focused on two basic design methodologies: sub-threshold current-mode circuits and above-threshold circuits. The former approach suffers from higher inhomogeneities (e.g. device mismatch) than the latter one, however it offers lower noise energy (noise power times bandwidth) and better energy efficiency (bandwidth over power)  \cite{LiuBook15}. Alternatively,   digital neurons can be realized using CMOS logic circuits such as adders, multipliers, and counters  \cite{Rajendran13}. From a circuit design point of view, the synapses modulate the input spikes and transform them into a charge that consequently create post-synaptic currents that are integrated at the membrane capacitance of the post-synaptic neuron \cite{LiuBook15}. The implementation of silicon synapses typically follows the mixed-signal methodology, though they can also be implemented in the digital domain by employing SRAM cells \cite{Rajendran13}.  Neurons and synapses can also be implemented using memristive technologies, as described in section \ref{devices}.

\section{State-of-the-art neuromorphic hardware}\label{chips}
In this section, we describe the salient features of state-of-the-art Si CMOS based neuromorphic chips. Although research in this domain is more than three decades old, starting with the pioneering work of Carver Mead \cite{Mead-02}, the discussion in this paper is limited to some of the recent demonstrations of large scale neuromorphic platforms that integrate more than 1000 neurons.

\subsection{SpiNNaker}
SpiNNaker is a digital system that has been designed to efficiently simulate large spiking networks, approaching the complexity of the brain, in real time \cite{SpiNNaker}. Its building blocks are ARM9 cores that can access a small amount of local memory, plus some additional memory that is shared across one multi-core chip. No global memory exists. Nodes can communicate via a high-throughput fabric that is optimized towards routing small messages (not larger than 72 bit) with high efficiency. SpiNNaker's event-driven design manifests itself in the message handling paradigm. A message arriving at a core triggers an interrupt that queues the packet for processing by that core. The system is optimized for small packet handler code that processes messages quickly, keeping queues short (i.e., not much larger than 1). SpiNNaker thus implements fundamental concepts of the brain, such as event-driven computation, locality of information, high fan-in and fan-out connectivity, and communication with tiny messages.

The SpiNNaker system is constructed from processor cores, 18 of which are grouped on a die in a chip. 48 such chips, with up to 864 cores (depending on manufacturing yield) are assembled on one board. Chips on one board communicate using a custom protocol using direct connections, wired in a hexagonal topology. Larger systems can be built by connecting 48-chip boards with fast serial interconnects. The largest system in existence today consists of 1 million processors, housed in ten 19-inch racks at the University of Manchester.

While SpiNNaker can be programmed directly, its software stack provides several levels of abstractions, with spiking network implementation being facilitated by PyNN. PyNN is a Python library that supports portability of network designs between a variety of neuronal simulators and hardware, including the BrainScaleS system (described below). SpiNNaker's PyNN interface provides several standard neuron models such as LIF and Izhikevich's dynamical systems model, along with common algorithms for synaptic plasticity, including STDP. 

The successor of this chip, named SpiNNaker 2, uses a more modern process technology, increases the number of cores per chip and adds linear algebra accelerators, and several other improvements. It has been used successfully for deep learning with sparse connectivity \cite{SpiNN2DeepR}.

\subsection{TrueNorth} 
TrueNorth is a million neuron digital CMOS chip that IBM demonstrated using  28-nm  process technology in 2014 \cite{truenorth}. The chip is configured as a tiled array of 4096 neurosynaptic cores, with each core containing $12.75\,$KB of local SRAM memory to store the synapse states, neuron states and parameters, destination addresses of fan-out neurons, and axonal delays. The digital neuron circuit in each core implements leaky-integrate-and-fire dynamics, and is time-multiplexed so as to emulate the operation of up to 256 neurons, which helps in amortizing the physical area and power budgets. Each core can support fan-in and fan-out of 256 or less, and this connectivity can be configured such that neurons in any core can communicate its spikes to one axon in any other core, and then to any   neuron  in that core. The spike-based communication and routing infrastructure also allow the integration of multiple chips; IBM has recently demonstrated the NS16e board that integrates 16 TrueNorth chips. The chip integrates 256 million SRAM synapses, with the synaptic connectivity programmable to two values - \{0,1\}, four programmable 9-bit signed integer weights per neuron, and programmable spike delivery time at the destination axon in the range of  $1-15$ time steps. The spike routing is completed asynchronously in every time step, chosen to be $1\,$ms, so as to achieve real-time operation akin to biology, although faster synchronization clocks permit accelerated network emulation. The corelet programming environment is used to map network parameters from software training to the TrueNorth processor.   Thanks to the event-driven custom design, the co-location of memory and processing units in each core, and the use of low-leakage silicon CMOS technology, TrueNorth can perform 46 billion synaptic operations per second (SOPS) per Watt for real-time operation, with $26\,$pico Joule per synaptic event; its  power density of $20\,$mW/cm$^2$ is about 3 orders of magnitude smaller than that of  typical CPUs.

\subsection{Loihi}
 Loihi is a neuromorphic learning chip developed by Intel using their 14-nm FinFET process in 2018 \cite{loihi}. This multi-core chip supports the implementation of axonal and synaptic delays, neuronal spiking threshold adaptation, and programmable synaptic learning rules based on spike timing and reward modulation. The chip has 128 neural cores, with each core having 1024 spiking neurons and $2\,$Mb SRAM to store the connectivity, configuration, and dynamic state of all neurons within the core. The chip also includes three embedded x86 processors, and  $16\,$MB of synaptic memory implemented in SRAM,   supporting synaptic bit resolution from 1 to 9 bits. Thus, it supports roughly 130,000 neurons and 130 million synapses. Spikes are transported between the cores in the chip using packetized messages by an asynchronous network-on-chip (NoC) and allows connecting to 4096 on-chip cores and up to 16,384 chips via hierarchical addressing. To address the scaling of network connectivity to biological levels (i.e., fan-out of 1000), Loihi supports several features including core-to-core multicast and population-based hierarchical connectivity. The cores in the chip can be programmed using microcodes to implement several forms of neuromorphic learning rules such as pairwise STDP, triplet STDP, certain reinforcement learning protocols, and other rules that depend on spike rates as well as spike-timing. At nominal operating conditions, Loihi delivers 30 billion synaptic operations per second, consuming about $15\,$pico Joule per synaptic operation. A python package for the Nengo neural simulator allows users to study the implementation of spiking networks on Loihi without accessing the hardware.

\subsection{BrainScaleS-1}
The BrainScaleS system is a mixed-signal platform that combines analog neuron circuits with digital communication \cite{Meier2015}. Its unique feature is a speedup-factor of $10^3$ to $10^4$ for spiking network emulations, meaning that a network model running for $1\,$s wall-time on the hardware emulates up to $10,000\,$s of ``biological'' simulation time. BrainScaleS supports the adaptive exponential integrate-and-fire model that can be parameterized to exhibit diverse firing patterns. The BrainScaleS system's smallest unit of silicon is the HiCANN (High-Input Count Analog Neuronal Network) chip \cite{schemmel10}. The number of neurons per chip can be configured within wide limits, following a trade-off between the number of input synapses and the number of neurons. A single chip supports a maximum of 512 spiking neurons and up to about 14,000 synapses per neuron. Larger networks are supported by wafer-scale integration of HiCANN chips, which are wired directly on the silicon wafer, without cutting it into discrete elements. A single wafer supports $4\times10^7$ synapses, and up to 180,000 neurons.

 Prototypes of the next BrainScaleS generation  support programmable plasticity via general purpose processors embedded on the die alongside the neuromorphic circuitry \cite{brainscalesHybridRISC}. These processors have access to dedicated sensors at the synapses that measure the time interval between pre- and post-synaptic spikes. Arbitrary functions can be defined that compute updates to synaptic weights from this information. This enables highly flexible learning rules to be implemented on BrainScaleS, including reward-based learning. BrainScaleS, like SpiNNaker, leverages the PyNN API (\url{http://neuralensemble.org/PyNN}) to allow the user to specify spiking networks for emulation on the hardware. 
 
\subsection{NeuroGrid/Braindrop}
The goal of the NeuroGrid platform is to implement large-scale neural models and to emulate their function in real time \cite{Y2014benjaminProcIEEE}. Hence, the system's memory and computing resources have time constants that are well matched to the signals that need to be processed. NeuroGrid employs analog/digital mixed-signal subthreshold circuits to model continuous time neural processing elements. The physics of field effect transistors operating in the subthreshold regime is used to directly emulate the various neuronal and synaptic functions. 

Neurogrid comprises of a board with 16 standard CMOS NeuroCore chips connected in a tree network, with each chip consisting of a 256$\times$256 array of two compartmental neurons. Each neuron in the array can target multiple destinations by virtue of an asynchronous multicast tree routing digital infrastructure. The full NeuroGrid board can implement models of cortical networks of up to one million neurons and billions of synaptic connections with sparse long-range connections and dense local connectivity. A mixed-signal neurosynaptic core called Braindrop, with 4096 neurons and 64 KB of weight memory has also been recently demonstrated, leveraging the variability of analog circuits for performing useful computations \cite{Neckar2019}.

\subsection{DYNAP}
Dynamic neuromorphic asynchronous processors (DYNAP) is a family of mixed-signal neuromorphic  chips from INI Zurich. DYNAP-SE is one such chip fabricated in $180\,$nm CMOS technology, integrating 1024 neurons and 64k synapses \cite{Moradi2018}. The chip is organized into 4 cores, with each core having 256 analog neurons. The temporal dynamics of neurons are implemented using ultra-low power (sub-threshold) analog circuits whereas the asynchronous digital circuits allow to program and re-program the network connectivity at run-time, enabling the configuration of recurrent networks, multi-layer networks, and any arbitrary network topology. The analog
circuits implement a wide range of features, including multiple types of synaptic and neural dynamics, including the spike-frequency adaptation mechanisms that have been recently shown to be crucial in implementing LSTM-like networks with spiking neurons. The asynchronous digital circuits implement a
hierarchical routing scheme that combines the best features of all previously proposed approaches (e.g. the tree-based scheme from NeuroGrid with the 2D-grid mesh scheme from SpiNNaKer) minimizing the memory requirements. Moreover, the device mismatch that is present in the analog circuits is exploited to implement neural sampling and reservoir-computing strategies that require variability.

\subsection{ODIN}
ODIN is a $28\,$nm digital neuromorphic chip demonstrated by Catholic University Louvain in 2019 supporting simple forms of on-chip spike-driven synaptic plasticity (SDSP)  \cite{Frenkel2019}. The core supports $256$ neurons  which can be configured to implement  first-order LIF dynamics as well as the second-order Izikevich dynamics. The neuronal parameters are stored in a $4\,$kB SRAM array, and a global controller is used to time-multiplex the neuron logic circuit to implement the dynamics of the neurons in a sequential fashion.  The core also integrates 3-bit $256^2$ synapses  which are implemented as a $32\,$KB SRAM array. An additional bit is used per synapse to enable or disable online learning. Using a subset of pre-processed images from the MNIST dataset, the chip demonstrated on-chip learning achieving $84.5\%$ accuracy, and consuming $15\,$nJ per inference with rank-order coding.

Table \ref{table:1}  summarizes the   state-of-the-art neuromorphic chips today, along with some of their key attributes. Note that representative numbers are reported in the table, and in some instances, it is possible to exceed the performance metrics quoted here by operating the chip at higher frequencies or other operating conditions.   Furthermore,  newer generation prototype chips that form the building block for the larger systems have been recently reported as alluded to in the text (especially for BrainScaleS and SpiNNaker), although we have included the specifications of the full-scale systems in this table. 

\begin{table}[h!]
\centering\caption{Comparison of state-of-the-art neuromorphic chips along with some performance attributes }\tiny
\begin{tabular}{|p{1.6cm}|p{1.9cm}|p{4cm}|p{5.5cm}|  }
 \hline
Chip & Technology&Integration density & Key functionality/performance metric \\ 
 \hline\hline
SpiNNaker & ARM968, $130\,$nm CMOS (next-gen prototypes: ARM M4F, $28\,$nm CMOS) & Up to 1 K neurons/core, 1 M cores. & Programmable numerical simulations with 72-bit messages, for real-time simulation of spiking networks\\ 

\hline
TrueNorth & Digital ASIC at $28\,$nm CMOS&1 M neurons, 256 M Synapses; 1-bit synaptic state to represent a connection, with 4 programmable 9-bit weights per neuron  &SNN emulation without on-chip learning;  $26\,$pJ per synaptic operation.\\

\hline
Loihi &Digital ASIC at $14\,$nm CMOS &130 k neurons, 130 M synapses with  variable weight resolution (1-9 bits)&Supports on-chip learning with plasticity rules such as Hebbian, pair-wise, and triplet-STDP, $23.6\,$pJ per synaptic  operation (at nominal operating conditions).\\%

\hline
BrainScaleS &Mixed signal waferscale system, $180\,$nm CMOS (next-gen prototype: $65\,$nm CMOS)&180 K neurons, 40 M  synapses per wafer & $10^3 -10^4$ fold acceleration of spiking network emulations, with hardware-supported synaptic plasticity. Next-gen prototype: programmable plasticity. \\ 

\hline
Braindrop &Mixed signal $28\,$nm CMOS  &4096 neurons, 64K programmable weights (with analog circuits that allow realization of all-to-all connectivity) &  $0.38\,$pJ per synaptic update, implements the single core of a planned million-neuron chip.\\

 \hline
DYNAP-SE & Mixed signal $180\,$nm CMOS&1024 neurons, 64K synapses (12-bit CAM)&Hybrid analog/digital circuits for emulating synapse and neuron dynamics, $17\,$pJ  per synaptic  operation\\

 \hline
ODIN & Digital ASIC at $28\,$nm CMOS&256 neurons, 64K synapses with $3$ bit weight and 1 bit to encode learning  &   $12.7\,$pJ per synaptic  operation, implements on-chip spike-driven plasticity     \\ \hline
\end{tabular}
\label{table:1}
\end{table}

\section{Neuromorphic computing with memristive devices}\label{devices}

\begin{figure}[ht!]
\centering
\includegraphics[width=0.8\columnwidth]{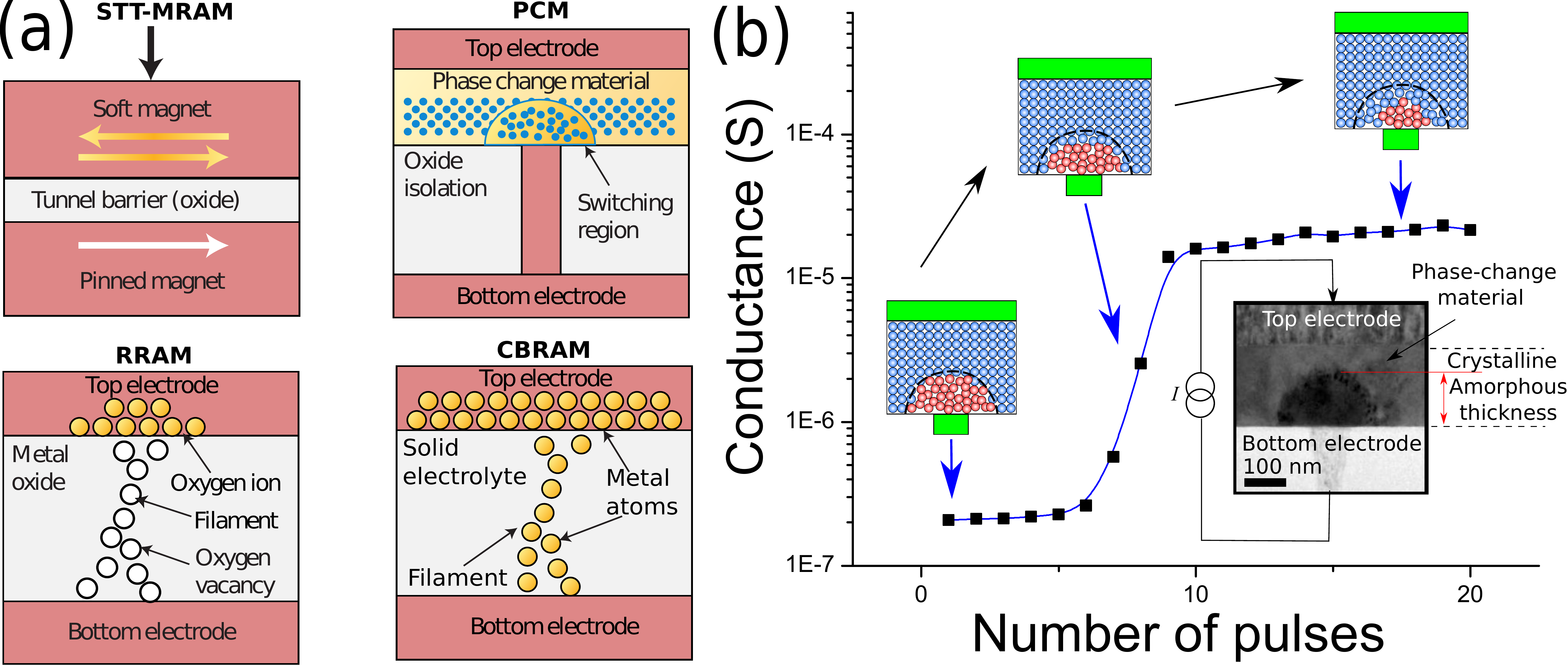}
\caption{(a) Schematic illustration of memristive devices STT-MRAM, PCM, RRAM, CBRAM. (adapted from \cite{Y2015wongNatNano}). (b) Incremental programming of PCM conductance (adapted from \cite{Y2016tumaNatNano}).} \label{fig:memdev}
\end{figure}

Going beyond conventional CMOS, a new class of emerging nanoscale devices, namely, resistive memory or memristive devices with their non-volatile storage capability, is particularly well suited for developing computing substrates for SNNs. In these devices, information is stored in their resistance or conductance states. The four main types of memristive devices are phase change memory (PCM), metal oxide based resistive random access memory (RRAM), conductive bridge RAM (CBRAM) and spin-transfer-torque magnetic RAM (STT-MRAM) (see Fig. \ref{fig:memdev}{a}). The resistance values of these devices are altered by the application of appropriate electrical pulses through various physical mechanisms, such as phase transition, ionic drift, and spintronic effects. Besides this ability to achieve multiple resistance values, many of these devices also exhibit an accumulative behavior whereby the resistance values can be incrementally increased or decreased upon the application of successive programming pulses of the same amplitude. These  attributes are key to their application in neuromorphic computing as illustrated in Fig. \ref{fig:memdev}{b} for PCM devices.

\begin{figure}[ht!]
\centering
\begin{tabular}{c}
\includegraphics[width=0.75\columnwidth]{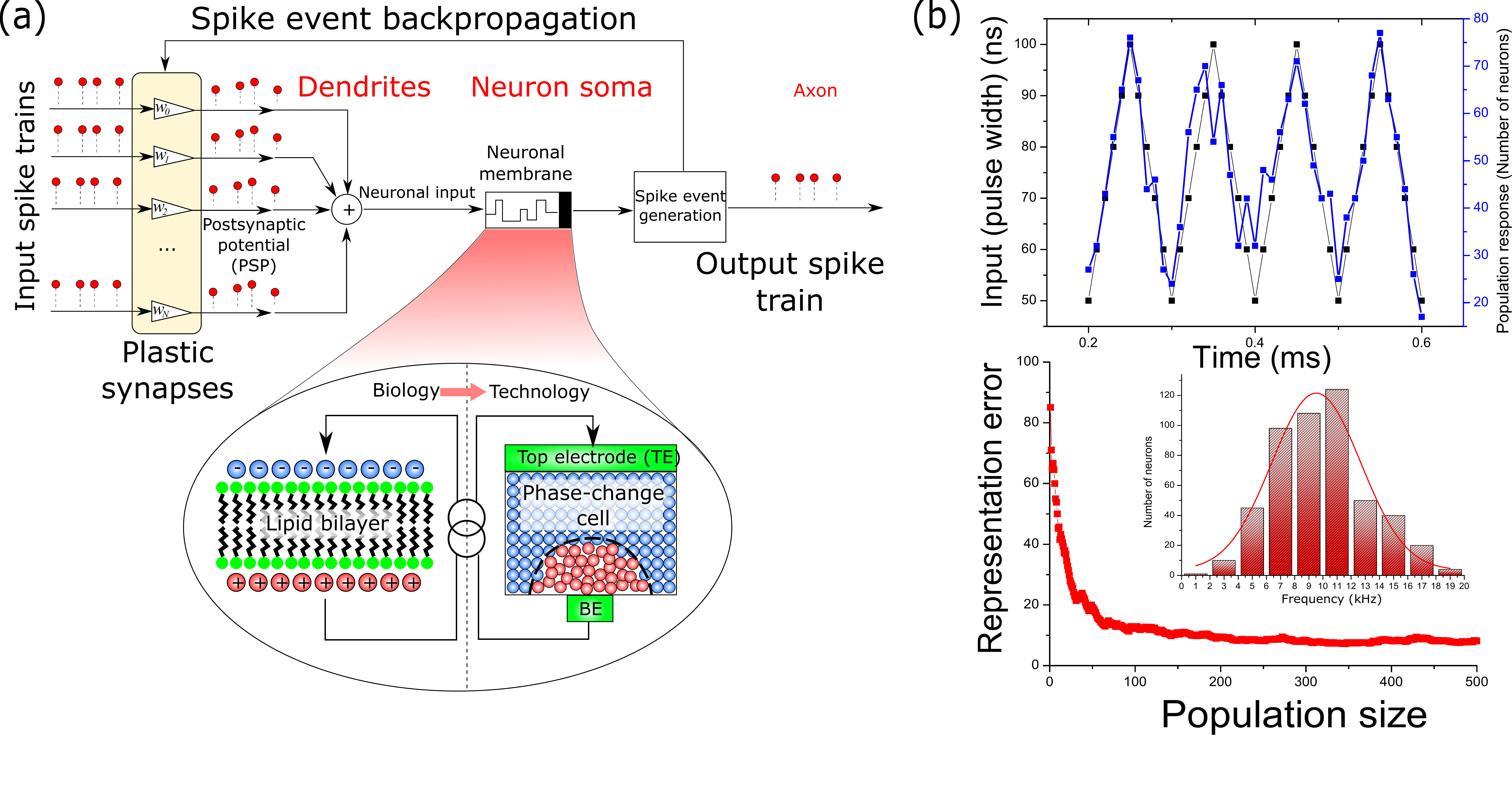}
\end{tabular}
 \caption{(a) Schematic illustration of phase-change neuron that consists of
the dendrites, the soma and the axon. The key element is the neuronal membrane that stores the membrane potential in the phase configuration of a phase-change memory device. It is possible to connect the dendrites to plastic synapses interfacing the neuron with other neurons in a network.  (b) The representation of high frequency signals via population coding of 500 slow-firing stochastic phase-change neurons. Also shown is the error in the representation of the stimulus by the population code.  The population code captures the input signal despite the fact that all the neurons in the population have their actual spiking frequency less than twice the base frequency of the input. Adapted from \cite{Y2016tumaNatNano}.}
\label{fig:memneuron}
\end{figure}

The accumulative behavior of memristive devices can be exploited to emulate neuronal dynamics \cite{Y2016tumaNatNano}. In one approach using PCM devices, the internal state of the neuron is represented by the phase configuration of the device (Fig. \ref{fig:memneuron}{a}). By translating the neuronal input to appropriate electrical signals, the firing frequency can be tuned in a highly controllable manner proportional to the strength of the input signal. In addition to the deterministic neuronal dynamics, stochastic neuronal dynamics also play a key role in signal encoding and transmission in biological neural networks. One notable example is the use of neuronal populations to represent and transmit sensory and motor signals. The PCM-based neurons exhibit significant inter-neuronal as well as intra-neuronal randomness, thus mimicking this stochastic neuronal behavior at the device level. Hence, multiple integrate-and-fire cycles in a single phase-change neuron could generate a distribution of inter-spike intervals and this enables population-based computation. By exploiting this, fast signals were shown to be accurately represented by population coding, despite the rather slow firing rate of the individual neurons (Fig. \ref{fig:memneuron}{b}).

\begin{figure}[ht!]
\centering
\includegraphics[width=0.8\columnwidth]{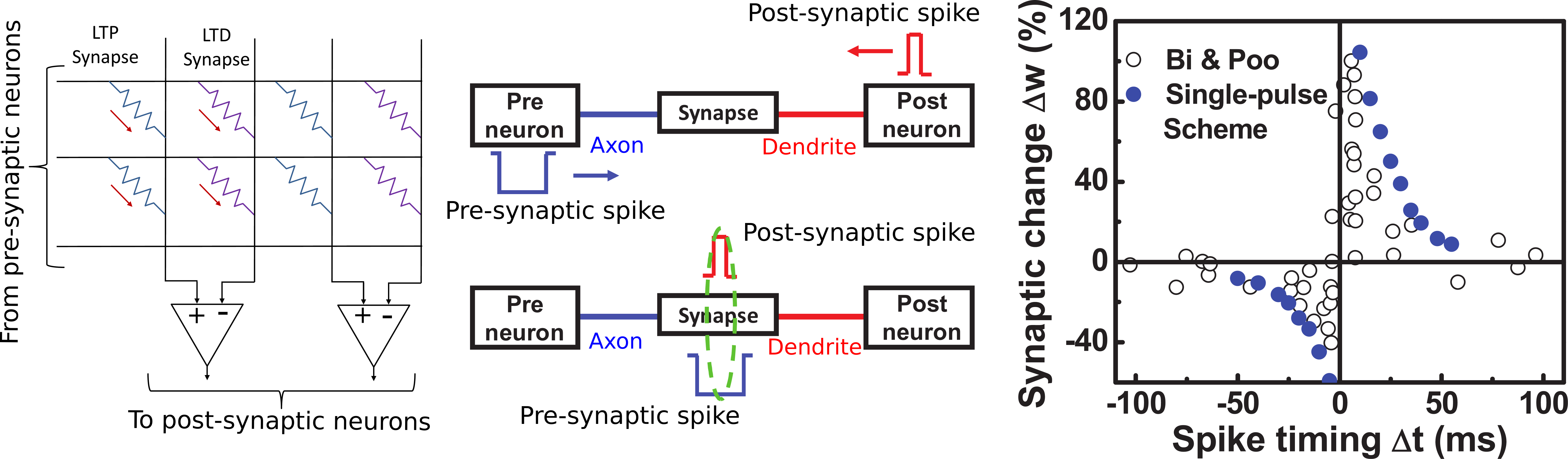}
\caption{Memristive devices organized in a crossbar configuration can be used to emulate synaptic communication and plasticity behaviors such as STDP by applying programming pulses from the periphery (adapted from \cite{Kuzum12}). Two devices, one for potentiation (LTP) and the other for depression (LTD) is used to represent the synapse in the array.} \label{fig:memsynapse}
\end{figure}

Memristive devices organized in a crossbar architecture can also be used to emulate the two essential synaptic attributes namely, synaptic efficacy and plasticity (Fig. \ref{fig:memsynapse}). Synaptic efficacy refers to the generation of a synaptic output based on the incoming neuronal activation; this can be realized using Ohm's law by measuring the current that flows through the device when an appropriate read voltage signal is applied. Synaptic plasticity, in contrast, is the ability of the synapse to change its weight, typically during the execution of a learning algorithm. The crossbar architecture is  well suited to implement synaptic  plasticity in a parallel and  efficient manner by the application of suitable  write pulses along the wires of the crossbar.

Although nanoscale devices offer exciting computational possibilities and scaling potential, several challenges  need  to be overcome to enable commercial products. PCM is based on the rapid and reversible phase transition of certain types of materials such as Ge$_2$Sb$_2$Te$_5$. However, it is necessary to reduce the programming current as well as improve the temporal stability of the achieved conductance states. RRAM and CBRAM typically rely on the formation and rupture of nanoscale filaments to achieve multiple conductance values. This filamentary mechanism is particularly prone to inter and intra-device variations which currently is the major technical hurdle. STT-MRAM devices consist of two magnetic layers separated by a tunnel barrier. These devices exhibit two resistive states depending on whether the magnetization of the two layers is in parallel or anti-parallel direction. These devices are expected to exhibit almost unlimited endurance, and faster switching   compared to RRAM and PCM. However, the key challenge is a substantially lower dynamic range in programmable conductance states (typically a factor of 2-3) as opposed to PCM and RRAM (which exhibit a dynamic range exceeding 100).  It is  crucial  that  new  circuits  and  architectures  are
developed that can mitigate these non-idealities  \cite{Y2018boybatNatComm}.
 There are also circuit-level challenges such as the voltage drop across the long wires connecting the devices as well as the overhead introduced by data converters and other peripheral circuitry. These aspects would limit the size of memristive crossbars that could be realized. However, in spite of these challenges, it is expected that we could achieve significant gains in areal and power efficiency by employing nanoscale memristive devices in neuromorphic processors  \cite{Rajendran15}.

\section{Signal processing applications}\label{apps}

Neuromorphic processors strive to balance the efficiency of computation with the energy needed for this computation, similar to the human brain. Systems enabled by such processors are expected to have the first impact on smart edge devices such as wearables, mobile devices, IoT sensors, and driver-less vehicles which have stringent constraints on size, weight, area, and power, and are required to intelligently interact with the world autonomously for extended periods of time. Neuromorphic approaches could result in highly power-efficient devices capable of responding quickly in an intelligent manner in dynamic environments for such applications.

\begin{figure*}[!ht]
	\centering
	\includegraphics[width=0.5\linewidth]{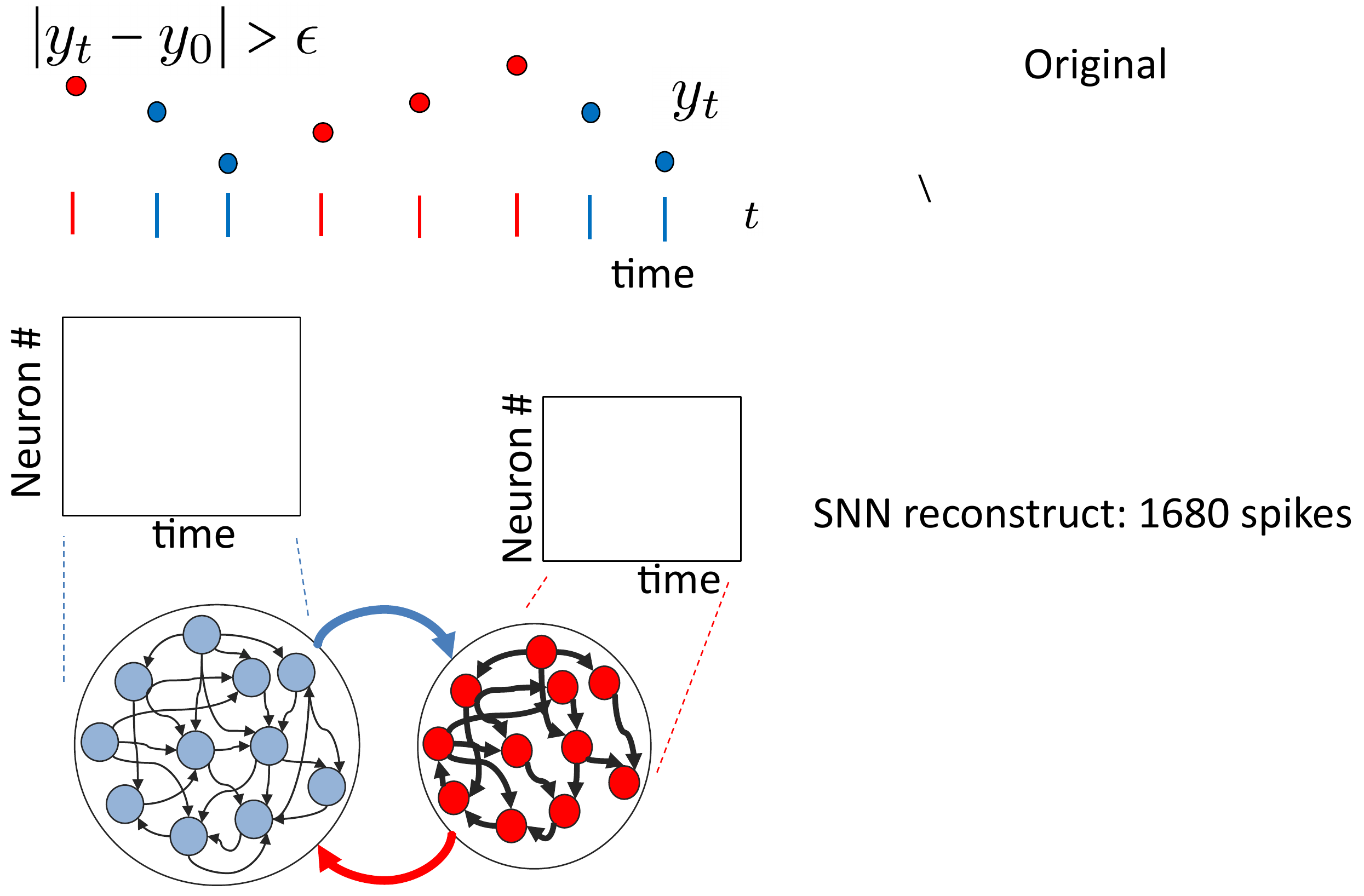}
	\caption{Sparse, event-triggered signal encoding which can be mapped efficiently to recurrent populations of spiking neurons with balanced spiking activity, i.e., between excitatory (blue) and inhibitory (red) neurons can enable efficient learning. In an illustrative example with this approach, fewer than 2000 spikes are needed to learn the features of an image that is conventionally   represented using more than 100,000 pixels. Adapted from \cite{nara3}. }	\label{fig:nara_02}
\end{figure*}
An illustrative example of this neuromorphic approach is the recent demonstration of a recurrent spiking neural network that learns directly from data streams while using only a limited number of training examples and a relatively small memory for learning \cite{nara2}.  Real-world data, represented in the form of spikes, was used for event-triggered learning which was implemented in a microprocessor operating in the asynchronous mode \cite{nara3}. SNNs naturally enable sparse representations of data and can be efficiently implemented using asynchronous logic because data processing occurs only during spike events (Fig. \ref{fig:nara_02}). By implementing local learning rules on networks with sparse connectivity, both the memory required to store the network parameters and the time needed to train the model can be minimized significantly compared to traditional machine learning approaches \cite{nara2}. This approach that implements SNNs in asynchronous processors has huge potential to enable edge devices that can learn and infer efficiently in the field.

Concomitant to the development of neuromorphic processors, bio-inspired event-driven sensors have also emerged, which can further accelerate the development of intelligent edge devices \cite{LiuBook15}. The most notable among them is the dynamic vision sensor (DVS) camera inspired by the information processing mechanisms of the retina and the silicon cochlea chip inspired by how the inner ear encodes sound signals in the spike domain. These sensors have superior performance compared to conventional sensors in several aspects.  For instance, the DVS camera has a $1000\times$ advantage in sampling rate   compared to a conventional camera, which helps in capturing fast changing events in the visual scene  \cite{Mueggler2014}. There are several demonstrations of systems that combine such event-driven sensors with efficient neuromorphic processors in recent years, some of which will be discussed below.

We now describe some of the proof-of-concept demonstrations targeting signal processing applications using the neuromorphic platforms discussed in section \ref{chips}. In a notable example using IBM's TrueNorth chip, deep neural networks were trained with a modified  backpropagation rule, so that the weights and neuronal dynamics can be easily ported to the hardware which supports only low precision synaptic weights; software equivalent performance was achieved for several benchmark pattern classification tasks with this approach \cite{Esser2016}. In another instance, a convolutional network running on TrueNorth that receives video input from a DVS camera was able to identify the onset of hand gestures with a latency of $105\,$ms while consuming less than $200\,$mW \cite{Amir2017}. Intel's Loihi has demonstrated over three orders of magnitude improvement in energy-delay product compared to conventional solvers running on a CPU for LASSO optimization problems by using a  spiking convolutional implementation of the Locally Competitive Algorithm \cite{loihi}.

Similarly, the SpiNNaker platform has already been used in several applications of spiking networks.  A large spiking model of visual cortex from a neural simulator running on a conventional compute cluster was recently ported to SpiNNaker, demonstrating comparable accuracy and favorable  speed and power consumption \cite{vanAlbadaSpinnvsNest}.  Several neuromorphic applications have also been pioneered on the BrainScaleS system, in particular on Spikey, the predecessor of the HiCANN chip. These include several networks performing various tasks from computational neuroscience \cite{Pfeil1311}, and the first published assessment of pattern recognition on   neuromorphic hardware \cite{Schmuker2014}. 

Due to its ability to operate in real time with biological neuronal signals, NeuroGrid has been used in a closed-loop brain-machine interface (BMI) application  \cite{Y2011dethierNIPS}.   A Kalman-filter based decoder was implemented via an SNN and tested in BMI experiments with a rhesus monkey. The success of this closed-loop decoder shows the promise of neuromorphic chips for implementing signal processing algorithms in a power efficient manner, which is a major enabling factor for the clinical translation of neural motor prostheses. The DYNAP-SE chip was recently used for reservoir computing that presented the first steps toward the design of a neuromorphic event-based neural
processing system that can be directly interfaced to surface EMG (sEMG) sensors for the on-line classification of the motor neuron output activities \cite{Y2018donatiIEEEBioCAS}.

\begin{figure}[h!]
\centering
\begin{tabular}{c}
\includegraphics[width=0.7\columnwidth]{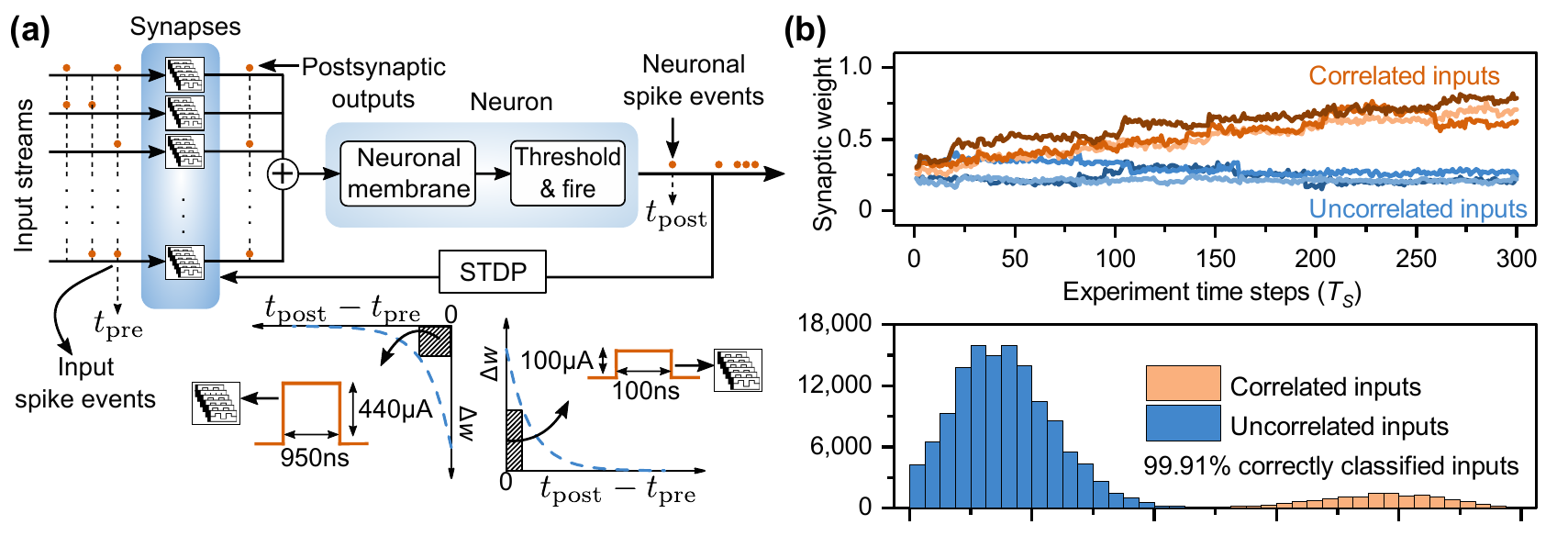}
\end{tabular}
\caption{(a) An SNN trained to perform the task of temporal correlation detection through unsupervised learning. (b) Synaptic weight
evolution as a function of time and the synaptic weight distribution at the end of the experiment. 10\% of the synapses receive correlated input data streams with a correlation coefficient of 0.75. Adapted from \cite{Y2018boybatNatComm}.} \label{fig:memcorr}
\end{figure}

In spite of being saddled with reliability and variability issues, memristive synapses have also demonstrated immense potential for signal processing applications. One noteworthy example was the demonstration of an SNN for detecting temporal correlations in an unsupervised fashion using plastic PCM synapses \cite{Y2018boybatNatComm}. (see Fig. \ref{fig:memcorr}). The network consists of a spiking neuron receiving an event-based data stream which is encoded as pre-synaptic input spikes arriving at PCM synapses; most of the data streams were temporally uncorrelated, but a small subset was chosen to be mutually correlated. Post-synaptic currents are generated at the synapses that received a spike and are integrated by the neuron which generates a spike when its membrane potential exceeds a threshold. An STDP rule was used to update the synaptic weights. Since the temporally correlated inputs are more  likely to eventually govern the neuronal firing events, the conductance of synapses receiving correlated inputs should increase, whereas that of synapses whose input are uncorrelated should decrease. Hence, the final steady-state distribution of the weights should show a separation between synapses receiving correlated and uncorrelated inputs. In the experiment, 144,000 input streams were fed through more than one million PCM devices representing the synapses. As shown in Fig. \ref{fig:memcorr}(b), well-separated synaptic distributions were achieved in the network at the end of the experiment, even though the devices exhibited significant device-to-device variability, and  drift in the programmed conductance states, demonstrating that nanoscale devices can enable complex computational capabilities in neuromorphic hardware platforms.

\section{Future Outlook}\label{Challenges}

It is widely believed that because of the added temporal dimension, SNNs should computationally be superior to, and thus transcend, the second generation deep neural networks. The energy-efficiency of neuromorphic systems based on SNNs makes them ideal candidates for embedded applications such as mobile phones, robotics, internet of things (IoT), and personalized medicine, which are subject to strict power and area constraints. Moreover, as Moore's law for CMOS scaling is coming to an end, these systems offer unique opportunities to leverage new materials and device structures going beyond standard CMOS processing. 

Going forward, there are algorithmic as well as technological challenges. From an algorithmic perspective, despite considerable advances, SNNs are yet to conclusively demonstrate superior performance compared to conventional deep learning, both in terms of accuracy and in many cases in terms of energy-efficiency as well. This gap in performance could be attributed to the lack of efficient and scalable supervised SNN learning algorithms, the lack of efficient local learning rules that can reach the performance of backpropagation, as well as the reliance on rate coding as opposed to more energy efficient temporal coding. Very recently, promising alternatives to rate coding that enable  efficient use of spike times have been introduced for SNNs \cite{Y2018kheradpishehNN}. Moreover, it was shown that recurrent SNNs with adaptive neurons can achieve classification performance comparable to state-of-the-art LSTM networks \cite{Bellec18}.
Furthermore,  efficient strategies have been demonstrated for converting deep neural networks to spiking networks for complex problems with negligible loss in accuracy \cite{Sengupta2019}.  Recently, a novel recurrent ANN unit called 
Spiking Neural Unit (SNU) was introduced that directly incorporates spiking neurons in deep learning architectures and achieves competitive performance by using backpropagation through time  \cite{Y2018wozniakArXiv}.  Finally, it should be noted that some types of neuromorphic hardware support deep learning directly (i.e., without spikes), and powerful algorithms have been developed to leverage the specific advantages of the architectures \cite{SpiNN2DeepR}. Also,  modifications of traditional deep learning algorithms have been proposed which enables their implementation in neuromorphic hardware using binary or ternary representations of neuronal activations and synaptic weights, although higher precision is required for gradient accumulation in these networks during training \cite{Courbariaux2015}. 

From a technology perspective, there are also numerous challenges associated with the use of memristive devices for neurmorphic computing. One key challenge applicable to all memristive technologies is related to the variations in the programmed conductance states with time and ambient temperature. The non-linearity and stochasticity associated with the accumulative behavior also pose scaling  challenges. Recent breakthroughs such as multi-cell architectures hold high promise to address these issues \cite{Y2018boybatNatComm}. 

In summary, we believe that there will be two stages of innovations for the field of low-power brain-inspired computing platforms. The near term innovations will come from neuromorphic accelerators built with conventional low-power mixed-signal CMOS architectures, which is  expected to lead to a period of transformative growth involving large neuromorphic platforms designed using ultra-low-power computational memories that leverage nanoscale memristive technologies. However, it has to be emphasized that the algorithmic exploration has to go hand-in-hand with advances in the hardware technologies.

{
\section{Authors and Bios}
\textit{Bipin Rajendran} is an Associate Professor of Electrical \& Computer Engineering at New Jersey Institute of Technology. Previously, he was a Master Inventor and Research Staff Member at IBM T. J. Watson Research Center in New York (2006-2012) and a faculty member in the Electrical Engineering  department at I.I.T. Bombay (2012-2015). His    research focuses   on   building   algorithms, devices,   and   systems   for brain-inspired  computing.    He  has  authored  over  70 papers  in  peer-reviewed  journals  and  conferences, and has been issued  59 U.S. patents. 

 \textit{Abu Sebastian} is a Principal Research Staff Member and Master Inventor at IBM Research – Zurich. He was a contributor to several key projects in the space of storage and memory technologies and currently leads the research effort on in-memory computing at IBM Research - Zurich. He is a co-recipient of the 2009 IEEE Control Systems Technology Award and the 2009 IEEE Transactions on Control Systems Technology Outstanding Paper Award. In 2015 he was awarded the European Research Council (ERC) consolidator grant. 

  \textit{Michael Schmuker} is a Reader in Data Science at the Department of Computer Science, University of Hertfordshire, UK. His research translates neurobiological principles of sensory computing into algorithms for data processing, inference and control, with a focus on neuromorphic olfaction and gas-based navigation. Michael has an MSc/Diploma in Biology, a PhD in Chemistry, and post-doctoral experience in Computational Neuroscience and Neuromorphic Computing. In 2014 Michael joined the University of Sussex on a Marie-Curie Fellowship (European Commission). In 2016 he joined the University of Hertfordshire as Senior Lecturer in Computer Science and promoted to Reader in Data Science in 2018.

\textit{Narayan Srinivasa} is currently the Director of Machine Intelligence Research Programs at Intel Labs. Prior to that, he was the Chief Technology Officer at Eta Compute focused on developing ultra low power AI solutions for audio applications.  From 2016-17,  he  was Chief Scientist and Senior Principal Engineer at Intel Labs, leading the work on the 14 nm Loihi neuromorphic chip. Prior to that, as the Director for the Center for Neural and Emergent Systems at HRL Laboratories in California, he focused on machine intelligence research and served as the principal investigator for the DARPA SyNAPSE, Physical Intelligence and UPSIDE programs. He  has  authored over 90 papers  in  peer-reviewed  journals  and  conferences, and has been issued  66 U.S. patents.

\textit{Evangelos S Eleftheriou} is currently responsible for the neuromorphic computing activities of IBM Research - Zurich. In 2002, he became a Fellow of the IEEE. He was also co-recipient of the 2003 IEEE Communications Society Leonard G. Abraham Prize and the 2005 Technology Award and of the Eduard Rhein Foundation. In 2009 he was a co-recipient of the IEEE Control Systems Technology Award and of the IEEE Transactions on Control Systems Technology Outstanding Paper Award. He was appointed an IBM Fellow in 2005 and was inducted into the US National Academy of Engineering (NAE) as Foreign Member in 2018.}
 
 \section{Acknowledgements}
BR was supported in part by the US National Science Foundation grant 1710009 and the grant 2717.001 from the Semiconductor
Research Corporation. MS received funding from the European Commission H2020 within the Human Brain Project, grant agreement ID 785907. AS acknowledges support from the European Research Council through the European Union’s Horizon 2020 Research and
Innovation Program under grant number 682675.

\ifCLASSOPTIONcaptionsoff
  \newpage
\fi

\bibliographystyle{IEEEtran}
\bibliography{refs}

\end{document}